\documentclass[showpacs,preprintnumbers,amsmath,amssymb,preprint]{revtex4}

\usepackage{epsfig}
\usepackage{graphicx}
\usepackage{dcolumn}
\usepackage{bm}
\usepackage{hyperref}
\usepackage{natbib}
\usepackage{color}

\begin{document}

\title{Consensus clustering approach to group brain connectivity matrices}
\author{Javier Rasero$^{1,2,3}$, Mario Pellicoro$^{2}$, Leonardo Angelini$^{2,3,4}$, Jesus M Cortes $^{1,5}$, Daniele
Marinazzo$^{6}$, and Sebastiano Stramaglia$^{2,3,4}$}

\affiliation{$^1$ Biocruces Health Research Institute. Hospital
Universitario de Cruces. E-48903, Barakaldo, Spain.}
\affiliation{$^2$ Dipartimento di Fisica, Universit\'a degli Studi
"Aldo Moro" Bari, Italy}
 \affiliation{$^3$ Istituto Nazionale di Fisica Nucleare, Sezione di
Bari, Italy} \affiliation{$^4$ TIRES-Center of Innovative
Technologies for Signal Detection and Processing, Universit\'a degli
Studi "Aldo Moro" Bari,  Italy} \affiliation{$^5$ Ikerbasque, The
Basque Foundation for Science, E-48011, Bilbao, Spain.
\\}
\affiliation{$^6$ Faculty of Psychology and Educational Sciences,
Department of Data Analysis, Ghent University, Henri Dunantlaan 1,
B-9000 Ghent, Belgium}
\date{\today}

\begin{abstract}
A novel approach rooted on the notion of {\it consensus} clustering,
a strategy developed for community detection in complex networks, is
proposed to cope with the heterogeneity that characterizes
connectivity matrices in health and disease.
The method can be summarized as follows: (i) define, for each
node, a distance matrix for the set of subjects by comparing the connectivity pattern of that node in all pairs of subjects (ii) cluster
the distance matrix for each node, (iii) build the consensus network
from the corresponding partitions and (iv) extract groups of
subjects by finding the communities of the consensus network thus
obtained. Differently from the previous implementations of consensus clustering, we thus propose to use the consensus strategy to combine the information arising from the connectivity patterns of each node. The proposed approach may be seen either as an exploratory technique or as an unsupervised pre-training step to help the subsequent construction of a supervised classifier. Applications on a toy model and two real data sets, show
the effectiveness of the proposed methodology, which represents
heterogeneity of a set of subjects in terms of a weighted network,
the consensus matrix. 
\end{abstract}

\maketitle In the  supervised analysis of human connectome data
\cite{Sporns_2011,Craddock_2013}, subjects are usually grouped under a common
umbrella corresponding to high-level clinical categories (e.g.,
patients and controls), and typical approaches aim at deducing a
decision function from the labeled training data, see e.g.
\cite{Fornito_2010}. However, the populations of subjects (healthy as well
as patients) is usually highly heterogeneous: clustering
algorithms find natural groupings in the data, and therefore
constitute a promising technique for disentangling the heterogeneity
that is inherent to many conditions, and to the cohort of controls.
Such an unsupervised classification may also be used  as a
preprocessing stage, so that the subsequent supervised analysis
might exploit the knowledge of the structure of data. Some studies
dealt with similar issues: semi-supervised clustering of imaging
data was considered in \cite{Filipovych_2011,Filipovych_2012}, other recent approaches cope
with the heterogeneity of subjects using multiplex biomarkers
techniques \cite{Steiner2016} and combinations of imaging and genetic
patterns \cite{Varol_2016}, whilst a strategy to overcome inter-subject
variability while predicting  behavioral variables from
imaging data has been proposed in \cite{Takerkart_2014}. Connectivity features have been used in data-driven approaches for analysis and classification of MRI data in \cite{Amico_2016,Iraji_2016}. The purpose of this
work is to introduce a novel approach that is rooted on the notion
of {\it consensus} clustering \cite{Lancichinetti_2012}, a strategy developed
for community detection in complex networks \cite{Barabasi_2002}.

To introduce our method, let us assume that a connectivity matrix is
associated to each item to be classified (usually a subject, but also individual scans for the same subject as in the example illustrated below). The goal of supervised analysis is to
mine those features of matrices which provide the best prediction of
 available environmental and phenotypic factors, such as task performance,
psychological traits, and disease states. When it comes to using
unsupervised analysis of matrices to find groups of subjects, the
most straightforward approach would be to extract a vector of
features from each connectivity matrix, and to cluster these vectors
using one of the commonly used clustering algorithms. The purpose of
the present work is to propose a new strategy for unsupervised
clustering of connectivity matrices. In the proposed approach the
different features, extracted from connectivity matrices, are not
combined in a single vector to feed the clustering algorithm;
rather, the information coming from the various features are
combined by constructing a {\it consensus} network \cite{Lancichinetti_2012}.
Consensus clustering is commonly used to generate stable results out
of a set of partitions delivered by different clustering algorithms
(and/or parameters) applied to the same data \cite{ISI:000183593700015}; here, instead, we use
the consensus strategy to combine the information about the data
structure arising from different features so as to summarize them
in a single consensus matrix.

\begin{figure}[ht!]
\begin{center}
\epsfig{file=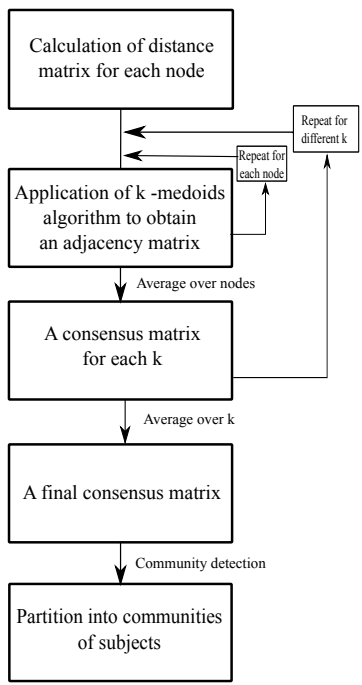,height=9.cm}
\end{center}
\caption{{\small The flowchart of the proposed methodology.
\label{fig1}}}
\end{figure}

The unsupervised strategy that we propose here to group subjects, without
using phenotypic measures, can be summarized as follows, and as depicted in figure
(\ref{fig1}): (i) define, for each node, of a distance matrix
for the set of subjects (ii) cluster the distance matrix for each
node, (iii) build the consensus network from the corresponding
partitions and (iv) extract groups of subjects by finding the
communities of the consensus network thus obtained . We remark that
the proposed approach not only provides a partition of subjects
in communities, but also the consensus matrix, which is a geometrical
representation of the set of subjects. In the next section we
describe in detail the method and apply it to a toy model, then we
show the application on two real MRI data sets. Finally, some
conclusions are drawn.

\section{Method}
Let us consider {\it m} subjects whose functional (structural) $N\times N$
connectivity matrix \cite{Rubinov_2010}, where N is the number of nodes,
will be denoted by $\{ \bf{A}(i,j)_\alpha\}$, $\alpha =1,\ldots,m$ and $i,j =1,\ldots,N$. For each node i, we build a distance matrix for the set of subjects as follows.
Consider a pair of subjects $\alpha$ and $\beta$, and consider the corresponding nodal connectivity patterns  $\{ \bf{A}(i,:)_\alpha\}$ and $\{ \bf{A}(i,:)_\beta\}$; let r be their Spearman correlation. As the distance between the two subjects, for the node i, we take $d_{\alpha \beta}=1-r$;  other choices for the distance can be used, like, e.g., $d_{\alpha \beta}=\sqrt{2(1-r)}$ where $r$ is the  Pearson correlation. The $m\times m$ distance matrix $d_{\alpha \beta}$
corresponding to node $i$  will be denoted by
$\bf{D}_i$, with $i=1,\ldots,N$. The set of $\bf{D}$ matrices may be
seen as corresponding to layers of a multilayer network
\cite{Boccaletti_2014}, each brain node providing a layer.

Each distance matrix $\bf{D}_i$ is then partitioned into $k$ groups of subjects using k-medoids method \cite{Brito_2007}. Subsequently, an $m\times m$ consensus
matrix $\bf{C}$ is evaluated: its entry $C_{\alpha \beta}$ indicates
the number of partitions in which subjects $\alpha$ and $\beta$ are
assigned to the same group, divided by the number of partitions N.
The number of clusters $k$ may be kept fixed, thus rendering the
consensus matrix depending on k; a better strategy, however, is to
average the consensus matrix over $k$ ranging in an interval, so as
to fuse, in the consensus matrix, information about  structures at
different resolutions.

The consensus matrix, obtained as explained before, is eventually
partitioned in communities by modularity maximization, with the consensus matrix $\bf{C}$ being compared against the ensemble of all consensus matrices one may obtain randomly and independently permuting the cluster labels obtained after applying the k-medoids algorithm to each of the set of distance matrices. More precisely, a modularity matrix is evaluated as $$\bf{B}=\bf{C}-\bf{P},$$
where $\bf{P}$ is the expected co-assignment matrix, uniform as a consequence of the null ensemble here chosen, obtained repeating many times the permutation of labels; the modularity matrix $\bf{B}$ is eventually submitted to a modularity optimization algorithm to obtain the output partition by the proposed approach (we used the Community Louvain routine in the Brain Connectivity Toolbox \cite{ISI:000280181800027}, which admits modularity matrices instead of connectivity matrices as input).

We remark that the proposed approach has similarities with
the one adopted in \cite{Shehzad_2014}, where techniques from genome-wide
association studies coping with the problem of a huge number of
comparisons were applied to connectomes, thus identifying nodes
whose whole-brain connectivity patterns vary significantly with a
phenotypic variable. The approach in \cite{Shehzad_2014} consists in two
steps. First, for each node in the connectome, a whole brain
functional connectivity map is evaluated, and then the similarity
between the connectivity maps of all possible pairings of
participants, using spatial correlation, is calculated. Then, in the
second stage, a statistics is evaluated for each node, indicating
the strength of the relationship between a phenotypic measure and
variations in its connectivity patterns across subjects. The main
similarity with the proposed approach is that in both methods, for
each node in the connectome, the comparison between the connectivity
maps yields a distance matrix in the space of subjects.

\section{A toy model}
As a toy model to describe the application of our method, we
simulate a set of 100 subjects, divided in four groups of 25 each. The
subjects are supposed to be described by 30 nodes. We will compare
our proposed approach with a standard procedure such as averaging
the distance matrices and then applying the clustering algorithm to
the average distance matrix.

The distance matrices corresponding to the first ten nodes are
constructed in the following way: the distance for pairs belonging
to the same group is sampled uniformly in the interval $[0.1, 0.4]$,
whilst the distance for pairs belonging to different groups is
sampled uniformly in the interval $[0.2, 0.4]$. The distance
matrices corresponding to the twenty remaining  nodes have all the
entries sampled uniformly in the interval $[0.2, 0.4]$. It follows
that in our toy model only 10 nodes, out of 30, carry information
about the presence of the four groups.

First of all, we evaluate the distance matrix among
subjects, averaged over the 30 nodes, and apply the k-medoids
algorithm to this matrix , searching for $k=4$ clusters (thus
exploiting the knowledge of the number of classes present in data);
this procedure leads to an accuracy of 0.89, measured as follows. Let us call $\{G_\alpha\}$, $\alpha =1,\ldots,4$ the four groups in the model and let M be the minimum between 4 and the number of clusters found by modularity maximization clustering; we denote $\{C_i\}$, $i =1,\ldots,M$ the largest M clusters found by clustering. The accuracy is then given by $$\frac{1}{m} \sum_{i=1}^M max_{\alpha} |G_\alpha \cap C_i|,$$ where $|G_\alpha \cap C_i|$ is the cardinality of the intersection of the two sets, and m=100 is the total number of subjects.

\begin{figure}[ht!]
\begin{center}
\epsfig{file=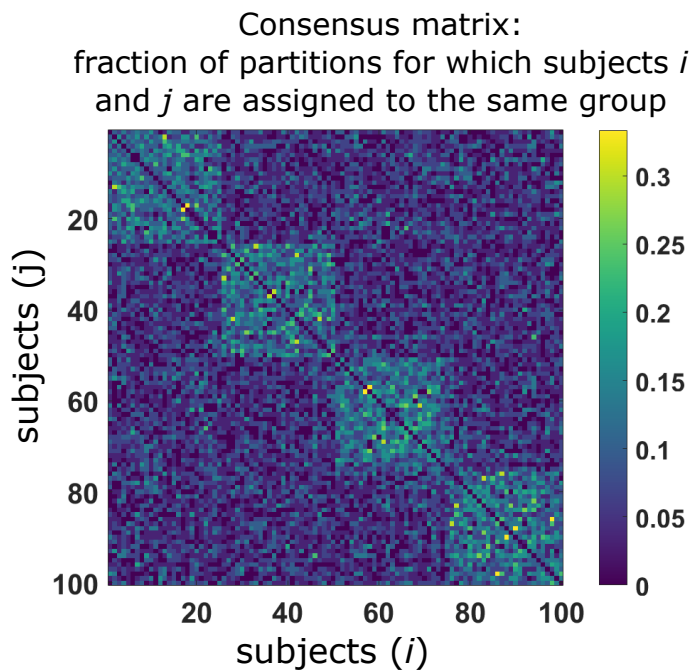,height=9.cm}
\end{center}
\caption{{\small Consensus matrix
among subjects in the toy model, obtained applying k-medoids with $k=10$ separately
to each of the 30 layers. Each entry $C_{\alpha \beta}$ of the matrix represents the number of partitions in which subjects $\alpha$ and $\beta$ were
assigned to the same group, divided by the number of partitions N
\label{fig2}}}
\end{figure}

Subsequently, we run the proposed approach by applying
separately to each distance matrix for each of the 30 nodes the
k-medoids algorithm with varying $k$. We then build the
corresponding consensus matrix. For example in figure (\ref{fig2})
the consensus matrix among subjects is depicted as obtained applying
k-medoids with $k=10$ separately to each of the 30 layers. Then, the
communities of the consensus matrices have been estimated as described in the previous Section.

\begin{figure}[ht!]
\begin{center}
\epsfig{file=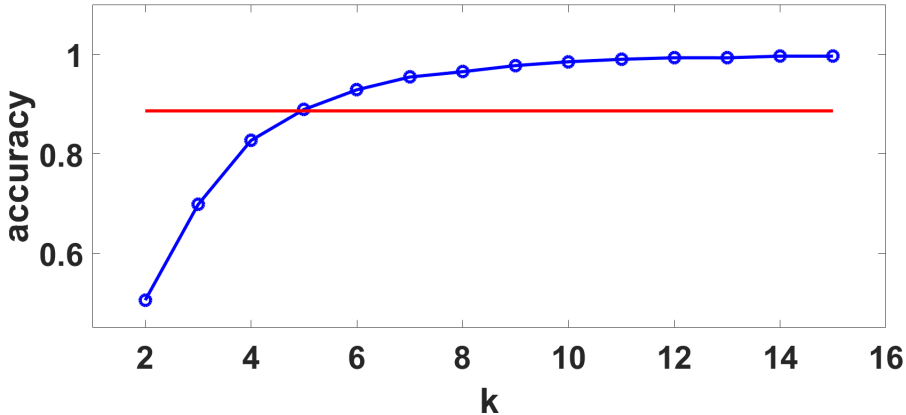,height=6.cm}
\end{center}
\caption{{\small The accuracy of the partition, provided by
modularity maximization on the consensus matrix, is depicted versus
$k$. The horizontal line represents the accuracy obtained by
clustering the average distance matrix using k-medoids and $k=4$.
\label{fig3}}}
\end{figure}

In figure (\ref{fig3}) the accuracy of the partition, provided by
modularity maximization on the consensus matrix, is depicted versus
$k$,  in order to show how it varies with $k$: 
it shows that the proposed method performs better than the
partition of the average distance matrix on this example, for large $k$; we remark that
the accuracy 0.89 is reached by k-medoids on the average distance
using $k=4$ i.e. exploiting the knowledge of the number of groups
present in the data set, whilst the proposed
algorithm determines both the number of clusters and the partition.
Intuitively, the proposed approach works better in this example for
large $k$, because in the distance matrix corresponding to an informative
node, due to chance, the block corresponding
to a group is seen as fragmented in smaller pieces; those pieces can
be retrieved using k-medoids with large $k$. On the other hand when
the consensus is made across the different informative nodes, all those pieces
merge in the consensus matrix and build the block corresponding to
the four groups.

It is also worth noting that the accuracy by clustering the
averaged consensus matrix (over the values of $k$) is one, i.e.
perfect group reconstruction. Averaging over the values of $k$
appears then to be a convenient strategy. Moreover, averaging over
values of parameters is a common strategy for consensus clustering,
hence building the consensus matrix while joining several values of
$k$ is in line with the philosophy of consensus clustering
\cite{Lancichinetti_2012}.

\begin{figure}[ht!]
\begin{center}
\epsfig{file=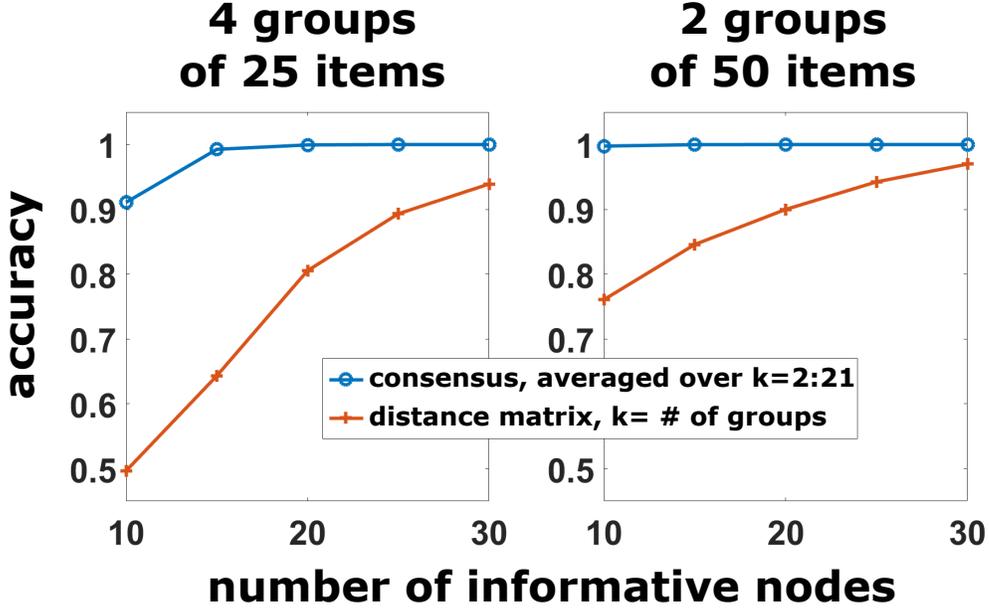,height=8.cm}
\end{center}
\caption{{\small The accuracy of the partition, provided by
modularity maximization on the consensus matrix averaged over twenty values of k, is depicted versus the number of informative nodes (when it is 30, all the nodes are informative). In the left panel the plots correspond to four groups of 25 subjects, the blue curve is the accuracy by the proposed method and the red line is the accuracy obtained by clustering the average distance matrix using k-medoids and $k=4$. In the right panel the case of two groups, each of 50 subjects, is considered; the blue line  is the accuracy by the proposed method and the red line is the accuracy obtained by clustering the average distance matrix using k-medoids and $k=2$. In all cases the consensus approach gives better results.
\label{fig4}}}
\end{figure}

 In order to show the effectiveness of the proposed approach under different conditions, we change the toy model by varying the number of informative nodes and the number of groups. We also use different parameters w.r.t. the previous simulations, the distance for pairs belonging
to the same group are still sampled uniformly in the interval $[0.1, 0.4]$, whilst the distance for pairs belonging to different groups is
sampled uniformly in the interval $[0.15, 0.4]$. The results, displayed in figure (\ref{fig4}), show that the proposed approach works better than the application of k-medoids to the average distance matrix.
\section{Application to real data sets}

\subsection{Longitudinal data set}
Growing interest is devoted to longitudinal phenotyping in
cognitive neuroscience: accordingly we consider here data from the
MyConnectome project \cite{Laumann_2015,Poldrack_2015}, where fMRI scans from a single subject were recorded over 18 months. In
\cite{Shine_2016} the presence of two distinct temporal states has been
identified, that fluctuated over the course of time. These
temporal states were associated with distinct patterns of
time-resolved blood oxygen level dependent (BOLD) connectivity
within individual scanning sessions and also related to significant
alterations in global efficiency of brain connectivity as well as
differences in self-reported attention. This data was obtained from the OpenfMRI database. Its accession number is ds000031.
The
functional MRI (fMRI) data was preprocessed with FSL (FMRIB Software
Library v5.0). The first 10 volumes were discarded for correction of
the magnetic saturation effect. The remaining volumes were corrected for motion, after which slice timing correction was applied to correct
for temporal alignment. All voxels were spatially smoothed with a
6mm FWHM isotropic Gaussian kernel and after intensity
normalization, a band pass filter was applied between 0.01 and 0.08
Hz. In addition, linear and quadratic trends were removed. We next
regressed out the motion time courses, the average CSF signal and the
average white matter signal. Global signal regression was not performed. Data were
transformed to the MNI152 template, such that a given voxel had a
volume of 3mm x 3 mm x 3mm. Finally we obtained 268 time series,
each corresponding to an anatomical region of interest (ROI), by
averaging the voxel signals according to the functional atlas described in
\cite{Shen_2013}.

Each of the 89 sessions resulted in a 268$\times$268 matrix of Pearson correlation coefficients.

\begin{figure}[ht!]
\begin{center}
\epsfig{file=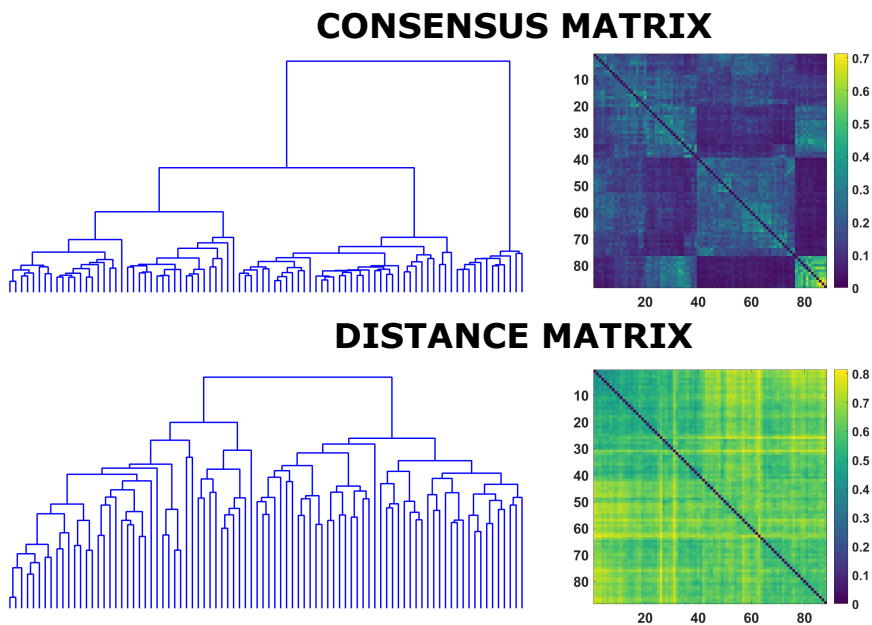,height=8.5cm}
\end{center}
\caption{{\small (Top) Concerning the {\it MyConnectome} data set, the
consensus matrix, obtained averaging over $k$, by the proposed
approach is displayed with nodes ordered according to hierarchical
clustering, with the corresponding dendrogram. (Bottom) The average
distance matrix, among the different sessions of the same subject,
and the corresponding dendrogram. \label{fig5}}}
\end{figure}

\begin{figure}[ht!]
\begin{center}
\epsfig{file=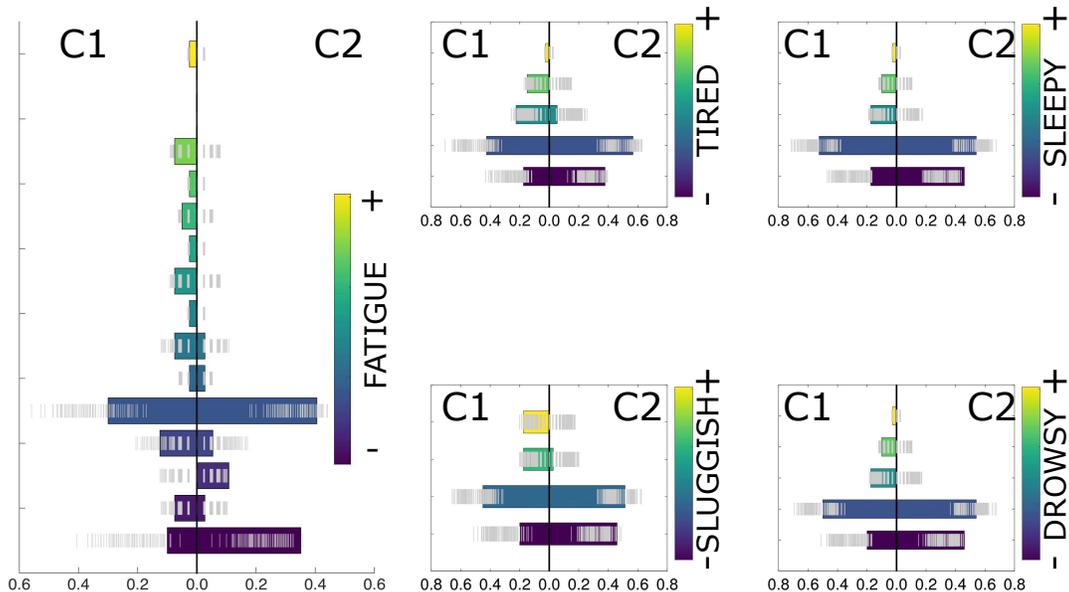,height=9.cm}
\end{center}
\caption{{\small {\it MyConnectome} data set: distributions of the values of the PANAS scores which are significantly different
among the two communities found by modularity optimization
on the consensus matrix provided by the proposed approach. An expected null distribution, whose quantiles are reported in gray, was obtained by shuffling the association between PANAS score and connectome matrix. \label{fig6}}}
\end{figure}

We treated the sessions as if they were connectivity matrices of
different subjects, and applied the proposed methodology.  In figure
(\ref{fig5}) we depict the distance matrix, among the different
sessions of the same subject, and the consensus matrix, obtained
averaging over ten values of $k$. Sessions are ordered, in both
cases, according to hierarchical clustering; the corresponding
dendrograms are also shown in the figure. It is clear that the
consensus matrix shows a hierarchical structure. Maximization of the
modularity provides two communities with modularity equal to 0.175. As
depicted in figure (\ref{fig6}), the two communities are significantly
different for several PANAS scores, all associated to tiredness. This is assessed visually using a null distribution obtained by shuffling 500 times the pairing between behavioral variable and connectome matrix and with a nonparametric Wilcoxon rank sum test: \textit{drowsy} (Bonferroni corrected p-value = $0.028$),  \textit{tired} (Bonferroni corrected p-value =
$0.041$), \textit{sluggish} (Bonferroni corrected p-value = $0.026$), \textit{sleepy}
Bonferroni corrected p-value = $0.012$), \textit{fatigue} (Bonferroni corrected p-value =
$0.022$). This confirms
the presence of two distinct temporal states. However the
hierarchical structure of the consensus matrix that we obtained
suggests that longer longitudinal recordings are needed to further 
evidence the richness of distinct functional states for single
subjects.

\begin{figure}[ht!]
\begin{center}
\epsfig{file=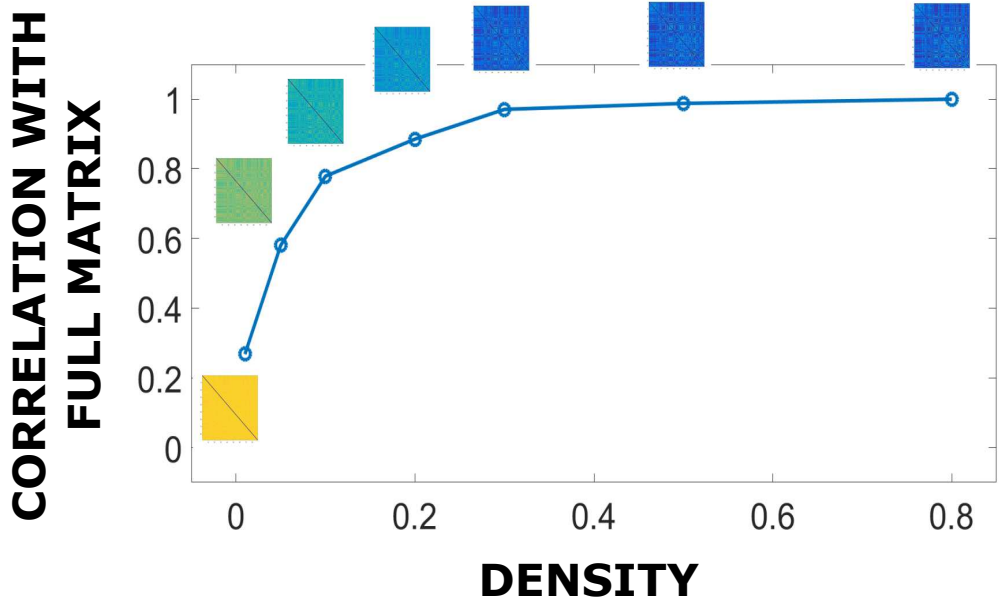,height=8.cm}
\end{center}
\caption{{\small The consensus matrix evaluated by the proposed approach, on the  brain connectivity matrices of the {\it MyConnectome} dataset, is compared with the consensus matrix from the proposed method on thresholded matrices. The linkwise similarity between the two consensus matrices is evaluated as the Pearson correlation of the corresponding entries in the two matrices, and is plotted versus the density of retained largest entries.  
\label{fig7}}}
\end{figure}

It is also worth considering the effects of network thresholding on the performance of the proposed algorithm: thresholding is a relevant problem in brain connectivity \cite{vanw_2010,devico_2017}. The functional networks in this data set are thresholded so as to retain a varying fraction (density) of the largest entries. In figure (\ref{fig7}) we plot the similarity between the consensus matrices obtained by the proposed algorithm after thresholding and the corresponding consensus matrix in the absence of thresholding, as a function of the density. The similarity between the consensus matrices is evaluated as the Pearson correlation between the entries of the two matrices. On one side the results show the robustness of the proposed approach to moderate thresholding, indeed up to  $20\%$ thresholding the consensus matrix is very close to what is obtained using the full matrices. On the other hand, the consensus matrix by the proposed approach is substantially different for sparser networks. This might speak to the fact that the correlation value is a debatable choice of a thresholding criterion for correlation matrices, and that the proposed approach is suited for weighted networks. 
\subsection{Resting healthy subjects, functional and structural connectivity}
\begin{figure}[ht!]
\begin{center}
\epsfig{file=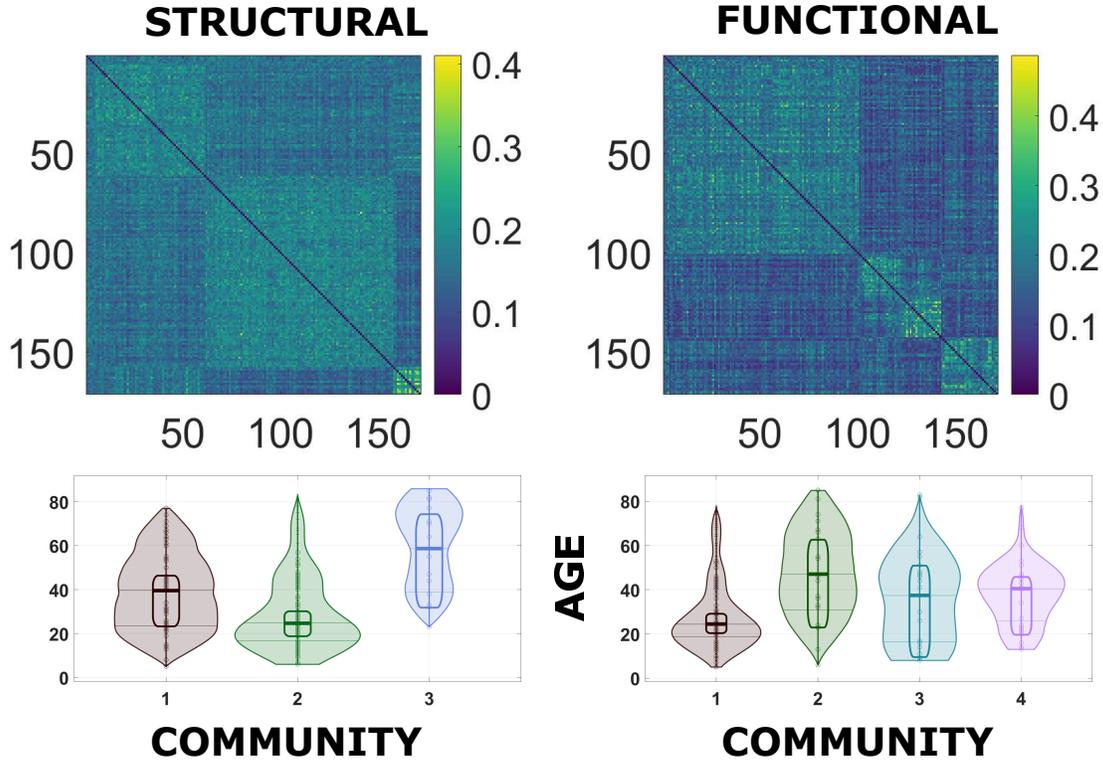,height=10.cm}
\end{center}
\caption{{\small (Top) Concerning the NKI data set, the consensus
matrices found by the proposed approach are shown for structural
(top-left) and functional (top-right) connectivity. (Bottom) The distribution of age values (in years) in the resulting communities are reported. The rectangles indicate the estimator with 95 percent high density interval, calculated by Bayesian bootstrap. The shaded areas indicate random average shifted histograms, with a kernel density estimate. The code for these plots is available at \url{https://github.com/CPernet/Robust_Statistical_Toolbox/}, courtesy of Cyril Pernet \label{fig8}}}
\end{figure}

We consider 171 healthy subjects from the NKI Rockland dataset
\cite{NKI}; for each subject we use both the structural Diffusion
Tensor Imaging DTI network and the functional network, already obtained from processed data as described in \cite{Brown_2012}. In this case the networks have 118 nodes. In figure (\ref{fig8}) we depict the
consensus matrix for both DTI and fMRI networks; modularity
maximization yields three communities for DTI networks and four communities for fMRI.
Concerning DTI, the three communities are significantly characterized by different age, with p-values equal to $9\times 10^{-4}$,  $2\times 10^{-5}$ and $0.003$ for the group comparisons 1-2, 2-3 and 1-3 respectively (see figure (\ref{fig8})). Considering fMRI data, the first group by the proposed algorithm have a different age than the second, the third and the fourth ones (taken as a whole) with probability $7\times 10^{-4}$. P-values here reported refer to a non-parametric ranksum test, similar significance was found using parametric tests.
We remark that our method   performs differently from  k-medoids over the average distance, where we obtain two groups with different age, t-test with probability $10^{-3}$ using the functional distance, whilst no significant difference in age using the structural connectivity.

Inspired by the results found by our method, we also performed a
multivariate distance regression \cite{Shehzad_2014}, that allowed us to build
a pseudo F-statistics to test whether age
correlates with the differences observed in the distance matrix for
each node. We have achieved this by comparing the observed
F-statistic with the pseudo F-distribution (that is not normal)
after $10^5$ data permutations. As expected, for both structural and
functional data, we found 124 and 76 nodes statistically related
with age respectively, thus suggesting that age is one of the
variables responsible of the community structure found by our
method.

\section{Conclusions}
An important issue such as dealing with the  heterogeneity that
characterizes healthy conditions, as well as diseases, requires the
development of effective methods capable to highlight the structure
of sets of subjects at varying resolutions. The approach that we
propose here is applied to sets of subjects each described by a
connectivity matrix; we propose a strategy, rooted in
complex networks theory, to obtain a consensus matrix which
describes the geometry of the data-set providing at different
resolutions groups of similar subjects. 
 Whilst the straightforward application of consensus clustering to a given data set combines the output from different clustering, our proposal, instead, is to apply a clustering algorithm separately to the connectivity map of each node. Hence the consensus strategy is exploited to combine the information arising from the different nodes.
Obviously, the choice of
k-medoids as the clustering algorithm for the individual layers is
not mandatory, other algorithms can be used, as well as the definition of the distance among subjects to be used by this algorithm. Moreover, in the present work the features that we considered are the
connectivity maps resulting from the whole brain connectivity pattern of each node, however other subsets of entries of matrices can be taken as well and the same strategy can be applied to fuse the different layers and produce a consensus matrix. Likewise, our framework is not limited to considering the whole brain and therefore it can be applied to analyze specific regions relevant to the problem at hand so as to exploit the benefits of our method.
Summarizing, our approach aims at disentangling the heterogeneity of groups corresponding to high-level categories, like healthy and disease,  finding natural groups within the cohort of patients (and within the cohort of controls). While dealing with data with both healthy and controls, it can be seen as a preprocessing step, that helps the subsequent construction of a supervised classifier healthy/subject.
\section*{Code}
The code for the construction of the consensus matrix, out of the set of connectivity matrices, is available at the
website https://github.com/jrasero/consensus

\section*{Acknowledgements}
The authors are grateful to Richard Betzel (University of Pennsylvania) and an anonymous referee for the most valuable suggestions. They also thank Guillaume Rousselet for valuable suggestions on data representation.

JR acknowledges financial support from the Minister of Education, Language Policy and Culture (Basque Government) under Doctoral Research Staff Improvement Programme.
We thank Guillaume Rousselet for valuable suggestions on data representation.

\bibliographystyle{ieeetr}
\bibliography{consensus}

\end{document}